\documentclass{aa}  

\usepackage{siunitx}
\usepackage{lscape}
\usepackage{ulem}
\usepackage{xcolor}
\usepackage{subfig}

\usepackage{graphicx}
\usepackage{txfonts}

\begin{document}

\title{New constraints on the presence of debris disks around \\
G\,196-3~B and VHS\,J125601.92$-$125723.9~b}

   \author{Olga V. Zakhozhay\inst{1,2}
          \and
         Mar\'{\i}a Rosa Zapatero Osorio\inst{3}
         \and
         V\'{\i}ctor J. S. B\'{e}jar\inst{4,5}
         \and
         Juan Bautista Climent\inst{6,8}
         \and
         Jos\'e Carlos Guirado\inst{6,7}
         \and
         Bartosz Gauza\inst{9}
         \and
         Nicolas Lodieu\inst{4,5}
         \and
         Dmitry A. Semenov\inst{1,10} 
        \and
        Miguel Perez-Torres\inst{11,12,13}
        \and
        Rebecca Azulay\inst{6,7}
        \and
        Rafael Rebolo\inst{4} 
        \and
        Jesús Martín-Pintado\inst{3}
        \and
        Charlène Lefèvre\inst{14}
          }

 \institute{Max-Planck-Institut f\"{u}r Astronomie,\ K\"{o}nigstuhl  17, 69117 Heidelberg, Germany\\ 
              \email{zakhozhay@mpia.de}
 \and
 Main Astronomical Observatory, National Academy of Sciences of the Ukraine, 03143 Kyiv, Ukraine
 \and 
Centro de Astrobiolog\'{\i}a, CSIC-INTA, Crta. Ajalvir km 4, E-28850 Torrej\'{o}n de Ardoz, Madrid, Spain\\ \vspace{-3mm}
 \and 
Instituto de Astrof\'{\i}sica de Canarias (IAC), Calle V\'{\i}a L\'{a}ctea s/n, E-38205 La Laguna, Tenerife, Spain\\ \vspace{-3mm}
 \and 
Universidad de La Laguna, Dpto. Astrof\'{\i}sica, E38206 La Laguna, Tenerife, Spain\\ \vspace{-3mm}
 \and 
Departament d’Astronomia i Astrofísica, Universitat de València, C. Dr. Moliner 50, E-46100 Burjassot, València, Spain\\ \vspace{-3mm}
 \and
Observatori Astronòmic, Universitat de València, Parc Científic, C. Catedrático José Beltrán 2, E-46980 Paterna, València, Spain\\ \vspace{-3mm}
 \and 
 Universidad Internacional de Valencia (VIU), C/ Pintor Sorolla 21, E-46002 Valencia, Spain\\ \vspace{-3mm}
 \and 
Janusz Gil Institute of Astronomy, University of Zielona G\'ora, Lubuska 2, 65-265 Zielona G\'ora, Poland\\ \vspace{-3mm}
 \and 
 Department of Chemistry, Ludwig Maximilian University, Butenandtstr. 5-13, 81377 Munich, Germany\\ \vspace{-3mm}
 \and 
Instituto de Astrof\'isica de Andaluc\'ia, Consejo Superior de Investigaciones Cient\'ificas (CSIC), Glorieta de la Astronom\'ia s/n, E-18008, Granada, Spain \\ \vspace{-3mm}
  \and
Facultad de Ciencias, Universidad de Zaragoza, Pedro Cerbuna 12, E-50009
Zaragoza, Spain\\ \vspace{-3mm}
  \and
School of Sciences, European University Cyprus, Diogenes street, Engomi,
1516 Nicosia, Cyprus\\ \vspace{-3mm}
  \and 
Institut de Radioastronomie Millim\'etrique (IRAM), 38406 Saint-Martin-d’H\'eres, France\\ \vspace{-3mm}
}

   \date{Received January 19, 2023; accepted March dd, yyyy}

 
  \abstract
   {The existence of warm (protoplanetary) disks around very young isolated planetary and brown dwarfs mass objects is known based on near- and mid-infrared flux excesses and millimeter observations. These disks may later evolve into debris disks/rings, although none has been observed/confirmed so far. Little is known about circum(sub)stellar and debris disks around substellar objects.}
   {We aim at investigating the presence of debris disks around two of the closest ($\sim$20 pc), young substellar companions, namely G\,196-3~B and VHS\,J125601.92$-$125723.9~b (VHS\,J1256$-$1257~b), whose masses straddle the borderline between planets and brown dwarfs. Both are companions at wide orbits ($\ge$100 au) of M-type dwarfs and their ages (50--100 Myr and 150--300 Myr, respectively) are thought to be adequate for the detection of second generation disks.}
   {We obtained deep images of G\,196-3~B and VHS\,J1256$-$1257~b with the NOrthern Extended Millimeter Array (NOEMA) at 1.3~mm. These data were combined with recently published Atacama Large Millimeter Array (ALMA) and Very Large Array (VLA) data of VHS\,J1256$-$1257~b at 0.87~mm and 0.9~cm, respectively.}
   {Neither G\,196-3~B nor VHS\,J1256$-$1257~b were detected in the NOEMA, ALMA and VLA data. At 1.3~mm, we imposed flux upper limits of 0.108 mJy (G\,196-3~B) and 0.153 mJy (VHS\,J1256$-$1257~b) with a 3-$\sigma$ confidence. Using the flux upper limits at the millimeter and radio wavelength regimes, we derived maximum values of 1.38$\times10^{-2}$~$M_{\rm Earth}$ and 5.46$\times10^{-3}$~$M_{\rm Earth}$ for the mass of any cold dust that might be surrounding G\,196-3~B and VHS\,J1256$-$1257~b, respectively. 
    }
   {We put our results in the context of other deep millimeter observations of free-floating and companion objects with substellar masses smaller than 20 M$_{\rm Jup}$ and ages between $\sim$1 and a few hundred million years.  
   Only two very young (2--5.4\,Myr) objects are detected out of a few tens. This implies that the disks around these very low-mass objects must have small masses, and possibly reduced sizes, in agreement with findings by other groups.  If debris disks around substellar objects scale down (in mass and size) in a similar manner as protoplanetary disks do, millimeter observations of moderately young brown dwarfs and planets must be at least two orders of magnitude deeper for being able to detect and characterize their surrounding debris disks.
   }

\maketitle
\section{Introduction} 

Substellar objects are known to host accretion and protoplanetary disks in a similar way as stars. First evidences of disks around brown dwarfs were given by the presence of strong and resolved H$\alpha$ emission line and near- and mid-infrared excesses \citep{Natta2001,Jaya2002,Jaya2003,Muzerolle2003,Muzerolle2005,Pascucci2003,Natta2004,Mohanty2005}.  The frequency of protoplanetary disks around brown dwarfs has been investigated in several star forming regions such as Taurus, IC\,348, Chamaeleon, $\sigma$ and $\lambda$ Orionis, $\rho$ Ophiuchus, Upper Scorpius, and TW Hya, all regions with less than 20 Myr, resulting in a rate ($\sim$40\,\%) very similar to that of low-mass stars  \citep{Luhman2005,Luhman2006,Luhman2010,Luhman2012,Guieu2007,Damjanov2007,Caballero2007,ZapateroOsorio2007,Hernandez2007,Barrado2007,Bouy2007,Scholz2007,Scholz2008,Alcala2008,Riaz2008,Riaz2009,Pena2012,Alves2013}. Spatially resolved images of protoplanetary disks surrounding massive brown dwarfs and proto-brown dwarfs in young star-forming regions have also been obtained thanks to the power of the Atacama Large Millimeter Array (ALMA) observations \citep{Ricci2014,Testi2016,Riaz2019,Rilinger2021}. These data indicate that the masses and sizes of young  circumsubstellar disks are a scaled version of the disks around low-mass stars.

Protoplanetary disks around brown dwarfs show evidence of growth toward planetesimals \citep{Apai2005,Ricci2014} and are expected to evolve into debris disks/rings. The time scale for the  protoplanetary--to--debris transition in the substellar regime is not well established so far and may depend from case to case. Preliminary studies suggest that brown dwarf disks transition to the debris phase by $\sim$40--50 Myr ages \citep{Riaz2014}. In the stellar regime, protoplanetary disks with large amounts of gas survive up to $\sim$10 Myr \citep[and references therein]{Wyatt2015}. Based on the high frequency of circumbsubstellar disks at very young ages, debris disks as descendants of the original disks around brown dwarfs should be common as well. But none is known/confirmed up to date to the best of our knowledge. Two candidates have been reported in the literature. V1400 Cen is a young, solar-mass star of the Sco-Cen OB association with an age of about 16 Myr. This star exhibited a remarkably long, deep, and complex eclipse event that was interpreted as it being eclipsed by a low-mass object (likely a brown dwarf or planet) orbited by a dense inner disk \citep{Mamajek2012,Kenworthy2015}. In a subsequent paper, \citet{Kenworthy2020} reported a non-detection of any flux in the Atacama Large Millimeter/submillimeter Array (ALMA) Band 7 with an upper limit 3\,$\sigma$ level of 57.6 $\mu$Jy at the position of the star. The authors concluded that the hypothesised bound ring system is composed of dust smaller than 1 mm in size, implying a young ring structure. The second candidate is G\,196-3\,B (L3 spectral type), the brown dwarf companion to a low-mass star with an age in the interval 50--100 Myr \citep{Rebolo1998}. \citet{Zakhozhay2017} argued that a feasible scenario to account for the very red colors of the L3 dwarf is the presence of a debris disk/ring. The lack of evidence of substellar debris disks is likely because many of the known brown dwarfs are old, lie at far distances, or remain unresolved from their more massive primaries. 

The aim of our current study is to investigate the possibility of circumplanetary or debris disk presence in two substellar companions that reunite unique conditions of youth, proximity, and companion--host separation: VHS\,J125601.92$-$125723.9\,b (from now on VHS\,J1256$-$1257\,b) and G\,196-3\,B. In this paper, we introduce the targets in Section~\ref{targets} and present the results of the 0.9~mm and 1.3~mm observations in Section~\ref{sec:observ}. The mass upper limits of any hypothetical cold dusty disk surrounding the targets are calculated in Section~\ref{sec:upperLimits}. Discussion and conclusions are presented in Sections~\ref{sec:discussion} and~\ref{sec:conclusions}.

\section{Target selection \label{targets}}

\par VHS\,J1256$-$1257 is a multiple system formed by at least three low-mass objects, which is exceptional for its proximity, 22.2~pc, and young age, 150--300~Myr, \citep{Gauza2015, Stone2016, Dupuy2020}. VHS\,J1256$-$1257 comprises a M7.5 primary component, which is actually an equal-magnitude close binary with a separation of 123~mas as revealed by the adaptive optics observations of \cite{Stone2016}, and one L7$\pm$1.5 companion that lies at a distance of $8"$ from the binary \citep{Gauza2015}. 
According to evolutionary models \citep{Chabrier2000}, both components of the binary have equal effective temperatures $T_{\rm eff}\approx 2620$~K~\citep{Stone2016} and masses of $M_{\rm \ast}=60-100$ M$_{\rm Jup}$ \citep{Dupuy2020}. The mass of the L7 companion is estimated in the interval 10--35 M$_{\rm Jup}$. In the work by \cite{Dupuy2020}, based on the trigonometric parallax of 22.2$^{+1.1}_{-1.2}$~pc{\footnote{The parallax was computed based on the observations obtained on the Canada-France-Hawaii Telescope (CFHT) in the period 2016--2019.}}, the authors derived a mass of 19$\pm 5$~M$_{\rm Jup}$ and a temperature of 1240$\pm 50$~K for VHS\,J1256$-$1257\,b. More recently, \cite{Dupuy2023}, using Keck adaptive optics imaging and aperture masking interferometry, have determined the primary binary orbit (a=1.96$\pm$0.03~AU, $P$ = 7.1$\pm$0.02~yr) and measured its dynamical total mass (0.141$\pm$0.008~M$_{\odot}$), consistent with being a brown dwarf or very low-mass stellar binary of 70--80~M$_{\rm Jup}$. The brightness, relatively wide orbit, and a lower contrast between the host star than in the case of other known directly imaged exoplanets, makes VHS\,J1256$-$1257\,b an exceptional object of which a thorough characterization is feasible. For example, intense photometric monitoring of this object has shown that this is the coolest object with larger photometric variability currently known \citep{Bowler2020,Zhou2022}. More recently, an intermediate resolution near-infrared spectrum of this object has been obtained with X-Shooter \citep{Petrus2023} and JWST has obtained a high fidelity 1-20 micron spectrum \citep{Miles2022}, which indicate the presence of water, methane, carbon monoxide, carbon dioxide, sodium, and potassium in its atmosphere, becoming a reference for exoplanet research. The spectral shape of VHS\,J1256$-$1257\,b seems to be influenced by disequilibrium chemistry and silicate clouds.

The {\sl Gaia} Early Data Release 3 (EDR3)~\citep{GaiaCollaboration2016,Lindegren2021} provides a distance of 21.15$\pm$0.21~pc for the primary binary component. This is the value we adopted for this paper. With this new determination and following the approach explained in \cite{Gauza2015}, the mass of VHS\,J1256$-$1257\,b turns out to be 20.5$^{+6.0}_{-4.5}$~M$_{\rm Jup}$, using the BT-Settl evolutionary models \citep{1998A&A...337..403B, 2003A&A...402..701B, Chabrier2000} with the \cite{2011SoPh..268..255C} solar abundances and the luminosity of log($L$/${\rm L_{\odot}}$) = -4.57$\pm$0.2 which we computed from the observed flux integration following the procedure described in \cite{Zakhozhay2017}. Uncertainty in the mass determination reflects the range of possible masses from the models given the age and luminosity in BT-Settl models. 
VHS\,J1256$-$1257\,b is confirmed to be young for its spectroscopic properties revealing a low-gravity atmosphere summarized by \citet{Gauza2015}. What is relevant for this work is the remarkable photometric property of VHS\,J1256$-$1257\,b: it is a very red source with near- and mid-infrared colors in excess with respect to dwarfs of related spectral types. Actually, VHS\,J1256$-$1257\,b is one of the reddest dwarfs known in the literature.

\par G\,196-3~B is the 15.3$^{+7.7}_{-2.7}$ M$_{\rm Jup}$, L3-type substellar companion of the nearby, high proper motion, M2.5 star G\,196-3\,A~\citep{Rebolo1998}. The system has a projected orbital size of 16$"$ at the parallactic distance of 21.81$\pm$0.01~pc~ \citep{GaiaCollaboration2016,Lindegren2021}. The youth of G\,196-3~B (50--100 Myr) is confirmed by the vast amount of optical and near-infrared spectroscopic observations: strong Li\,{\sc i} and weak Na\,{\sc i} and K\,{\sc i} absorption, strong VO and H$_2$O bands, plus the peaked $H$-band continuum \citep{Martin1999,Kirkpatrick2001,McGovern2004,McLean2007,Allers2007,ZapateroOsorio2014}. As VHS\,J1256$-$1257\,b, G\,196-3~B, with $T_{\rm eff} \approx 1850$ K, also has very red colors that clearly deviate from the indices of field dwarfs of similar spectral classification \citep{ZapateroOsorio2010}. This reddening behavior is not observed in younger (lower gravity) objects of related temperatures \citep{Lodieu2013,Pena2012,ZapateroOsorio2017}. Alternatively, \cite{Zakhozhay2017} studied the physical  feasibility of a debris disk being responsible for the infrared flux excesses. The authors modeled the spectral energy distribution of G\,196-3B from the optical through 24 $\mu$m and found that the best-fit solution is provided by a combination of an unreddened L3 atmosphere and a warm ($\sim$\,1300\,K), optically thick, narrow debris disk (``ring'') containing sub-micron particles and a mass $\ge 7 \times 10^{-10}$ M$_{\rm earth}$ located very near the central substellar object. Such a disk may resemble the rings of Neptune and Jupiter, except for its high temperature.

\section{Observations and data reduction}
\label{sec:observ}
\subsection{NOEMA}
We observed both systems VHS\,J1256$-$1257 and G\,196-3~B using ten 15-m antennas of the NOrthern Extended Millimeter Array (NOEMA). NOEMA is currently the most advanced millimeter interferometer in the Northern Hemisphere situated at 2550 m on the Plateau de Bure in the French Alps \footnote{\url{https://www.iram-institute.org/EN/content-page-56-7-56-0-0-0.html}}. The interferometer is equipped with a high-performance wide-band correlator named PolyFiX\footnote{More details about NOEMA can be found here: \url{https://www.iram.fr/IRAMFR/GILDAS/doc/html/noema-intro-html/noema-intro.html}}. 

All antennas are equipped with 2SB receivers, which provide low noise performance and long-term stability.
Observations were performed on 2019 March 25 and 27 for VHS\,J1256$-$1256 and G\,196-3~B, respectively. We used NOEMA in compact configuration D, with baselines between 24 and 176 m. This configuration allowed us to obtain images at 1.3~mm (230 GHz) with an angular resolution of 2\farcs53\,$\times$\,1\farcs43 at $0^{\circ}$ for VHS\,J1256$-$1257 and 1\farcs73\,$\times$\,1\farcs33 at $64^{\circ}$ for G\,196-3~B. The NOEMA field of view of 21\farcs9 at 1.3~mm was centered at the middle distance between the primary and the companion of VHS\,J1256$-$1257 and at the substellar companion G\,196-3~B (the primary G\,196-3~A was outside of the field of view). The total on-source integration time was 1.5 h for VHS\,J1256$-$1257 and 2.1~h for G\,196-3~B. 

Data calibration was performed with the GILDAS-CLIC software (sep-2019 version)\footnote{\url{http://www.iram.fr/IRAMFR/GILDAS/}}. Continuum was obtained by averaging line-free channels over the 7.744 MHz width (USB) centered at 230.0 GHz. We obtained a root mean square (rms) noise of 0.051 mJy\,beam$^{-1}$ (1-$\sigma$) for VHS\,J1256$-$1257~b and 0.036 mJy\,beam$^{-1}$ (1-$\sigma$) for G\,196-3~B in the synthetized beam, corresponding to minimum detected signals of 0.153~mJy and 0.108~mJy (adopting a  3-$\sigma$ detection limit). Images were cleaned using robust weighting with the H\"ogbom's method \citep{hogbom2003} and a robust parameter of 3. To obtain the convergence of the flux 1700 iterations were used. Nevertheless, even by tapering baselines above 50~m projected length, no detections were found. 

\subsection{ALMA}
ALMA observations were obtained for the VHS\,J1256$-$1257 system on 2019 March 07 between 5:22 and 07:27 UT using 43 of the ALMA 12 m antennas in Band 7 or $\sim$0.9 mm (G\,196-3 is not observable from ALMA latitude). The main scientific objective was to detect the continuum emission from a possible disk around the companion VHS\,J1256$-$1257~b. The central binary VHS\,J1256$-$1257 AB was observed simultaneously as it lies at the north-western border of the ALMA field of view. Two ALMA execution blocks were carried out, yielding a total on-source time of 73 min. The configuration was nominally TM1\@. The longest baseline was 313 m long, while the shortest was 15 m.
This resulted in a synthetized beam of 0\farcs91\,$\times$\,0\farcs81 with a position angle (P.A.) of $-$84.8$^{\circ}$. The precipitable water vapor (PWV) in the atmosphere above ALMA was between 1.21 mm and 1.26 mm during the observations. We used a full continuum sensitivity configuration. However, we adopted a mixed spectral (line and continuum) setup to maximize the potential scientific outcome of the project by also measuring the CO 3-2 line at 345.8 GHz in base band 1 with a spectral resolution of 0.98 km\,s$^{-1}$ covering a bandwidth of 1.875 GHz (1920 channels). Base band 2 was centered 1.8 GHz below the central frequency (344.0 GHz) to cover 2 GHz at a lower resolution (27.2 km\,s$^{-1}$; 128 channels). Likewise, the other two base bands were centered at the standard image band frequencies (357.8 and 356.0 GHz, respectively). Flux, band-pass, and phase calibrators were observed following the standard ALMA calibration procedure; VHS\,J1256$-$0547 served as band pass and flux calibrator, while J1305$-$1033 was used as phase calibrator. 

The calibration of the ALMA data followed the standard ALMA Quality Assurance procedure for Cycle 6 using the calibration pipeline 42254 (ALMA pipeline team\footnote{ALMA Pipeline Team, 2017, ALMA Science Pipeline User's Guide, ALMA Doc 6.13}) based on the CASA data analysis package version 5.4.$-$70 (McMullin et al.\ 2007). The calibrated data were then imaged with the \textit{tclean} task of the same CASA package in mfs mode, that is, combining all spectral channels into one image. VHS\,J1256$-$1257 data were imaged as a single field with a pixel size of 0\farcs16 and natural weighting in order to optimize the point-source sensitivity. No significant sources were found in the final image, resulting in an rms noise in the central region of 0.02~mJy\,beam$^{-1}$ (1-$\sigma$).

The same NOEMA and ALMA data of VHS\,J1256$-$1257 were presented in \citet{Climent2022}, where the peculiar emission of the primary component was examined in detail. Here, we focus on the very low-mass companions VHS\,J1256$-$1257~b and G\,196-3 B. To complement the NOEMA and ALMA data of VHS\,J1256$-$1257~b, we also used the {\sl Karl G. Jansky} Very Large Array (VLA) data at 6 GHz and 33 GHz (5 and 0.9 cm) of \citet{Climent2022}. While the (unresolved) primary component was clearly detected at both frequencies, the companion VHS\,J1256$-$1257~b remained below the sensitivity of the final VLA images with a 1-$\sigma$ rms noise of 0.004 and 0.010 mJy\,beam$^{-1}$ at 6 GHz and 33 GHz.

\begin{figure*}[!ht]
\begin{center}
\subfloat[]{\label{fig:vhs1257b}\includegraphics[width=0.5\textwidth]{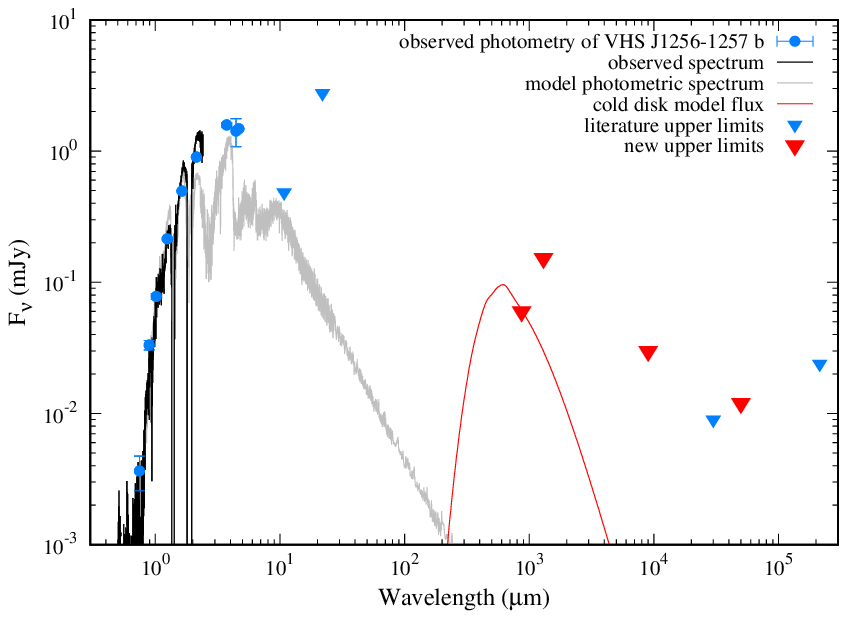}}
\subfloat[]{\label{fig:g196-3B}\includegraphics[width=0.5\textwidth]{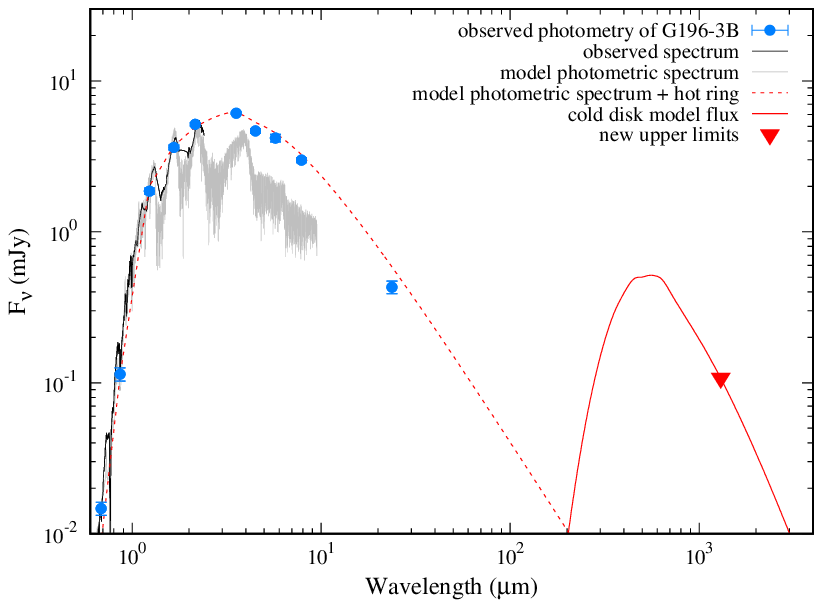}}
\caption{
The observed spectral energy distributions (SEDs) of the substellar companions VHS\,J1256$-$1257~b~(left panel) and G\,196-3~B~(right panel) are shown with blue circles (measured photometry), triangles (flux upper limits), and black lines (observed optical and near-infrared spectra). The new 3-$\sigma$ flux upper limits on the NOEMA, ALMA and VLA data reported here are displayed in red color. The theoretical BT-Settl  photospheric spectrum corresponding to the $T_{\rm eff}$ of each target is shown with the gray, solid line \citep{Allard2003,Allard2012}. In the right panel, we also illustrate with the dotted line the best model (photosphere and the hot ring) derived for G\,196-3~B by \citet{Zakhozhay2017}. The red solid line stands for the SEDs of putative cold dusty disks computed as explained in the text. 
}
\label{fig:SEDs}
\end{center}
\end{figure*}

\begin{figure*}
\begin{center}
\subfloat[]{\label{fig:upperlimitsMz_BD}\includegraphics[width=0.5\textwidth]{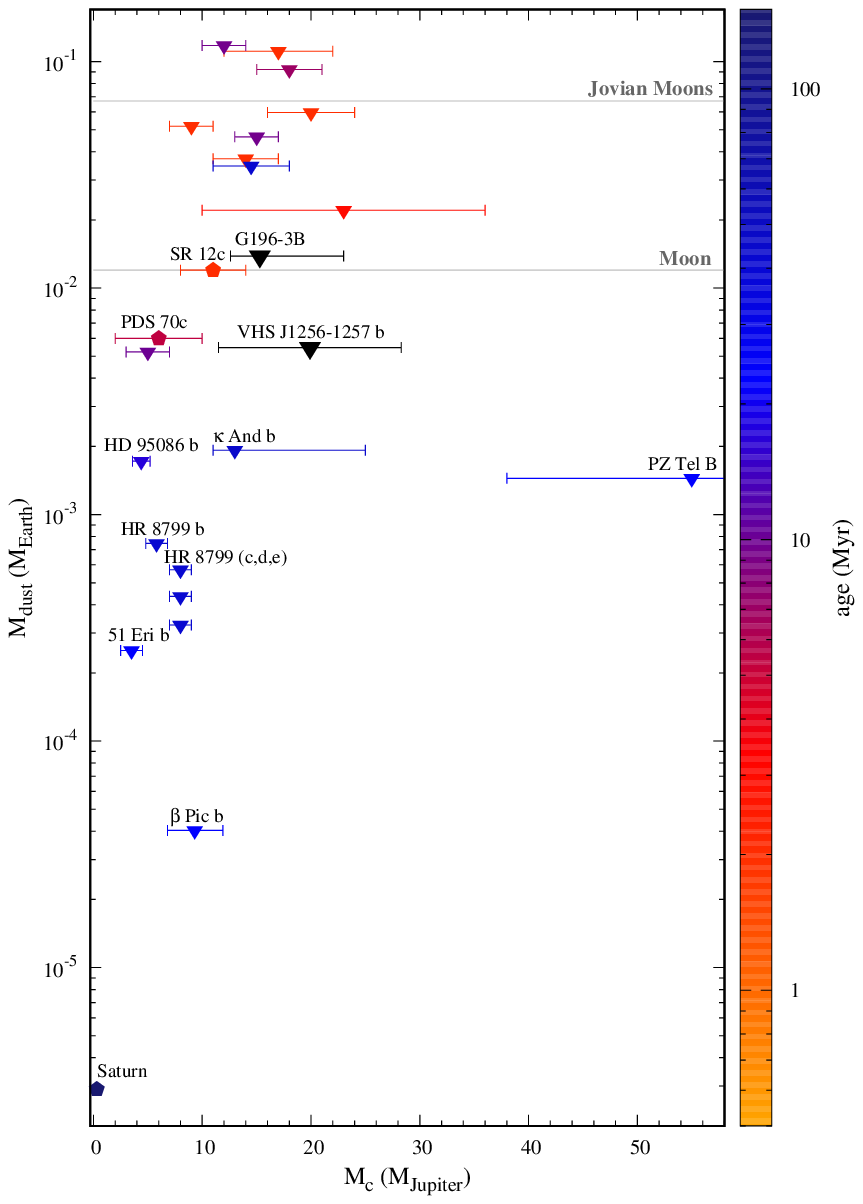}}
\subfloat[]{\label{fig:upperlimitsMc_BD}\includegraphics[width=0.5\textwidth]{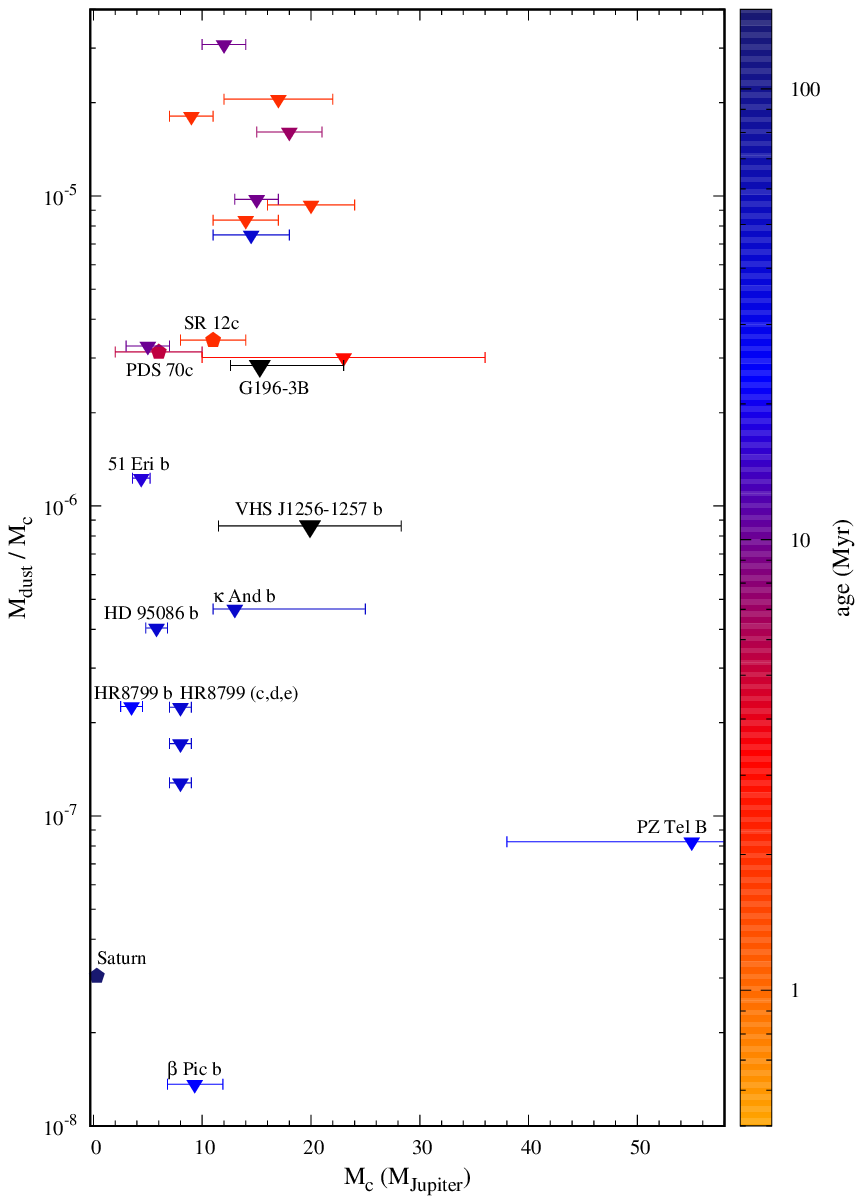}}
\caption{
 Upper limits (triangles) on the cold disk dust mass as a function of the mass of the companion for all substellar companions below $\sim$20 M$_{\rm Jup}$ and with sub-mm and/or mm data published in the literature. The age of the systems is color-coded as illustrated by the color bar, except for our targets VHS\,J1256$-$1257~b (150--300 Myr) and G\,196-3~B (50--100 Myr), which are shown with the black triangles. The left panel shows the upper limits on the dust mass in Earth mass units (${\rm M_{Earth}}$), while the panel on the right shows the masses of the disks relative to the masses of the substellar hosts. Grey horizontal solid lines in the left panel depict the masses of the Moon and of the Jovian Moons. The two confirmed circumsubstellar protoplanetary disks around SR~12~c and PDS~70 c are shown with hexagons. The disk dust masses are computed using the approach described in Section~\ref{sec:upperLimits}. Several objects, some of which are mentioned in the text, are labeled.
}
\label{fig:upperLimits_comanions}
\end{center}
\end{figure*}

\begin{figure}[!ht]
\begin{center}
\subfloat[]{\label{fig:vhs1257b}\includegraphics[width=0.5\textwidth]{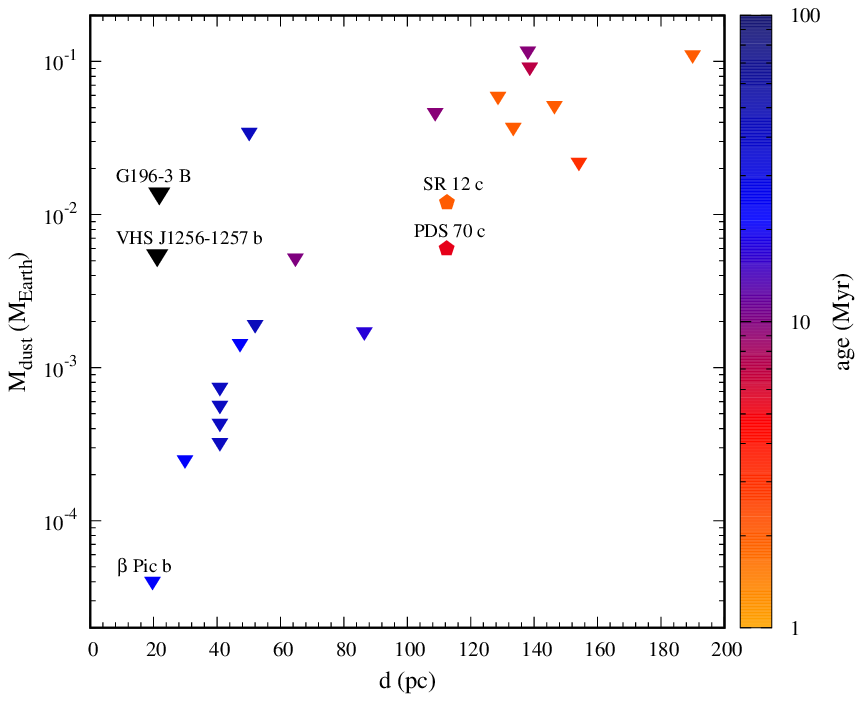}}

\caption{
Upper limits on the dust mass (triangles) versus distance for all objects of Table~\ref{tab:properties_literature}. Colored pentagons stand for the two known protoplanetary disks around SR~12~c and PDS~70 c. The age of the systems is color-coded as illustrated by the color bar, except for our targets VHS\,J1256$-$1257~b (150--300 Myr) and G\,196-3~B (50--100 Myr), which are shown with the black triangles. The disk dust masses are computed using the approach described in Section~\ref{sec:upperLimits}.
}
\label{fig:Md_d_age}
\end{center}
\end{figure}

\section{Dusty disk mass upper limits}
\label{sec:upperLimits}

\par Neither VHS\,J1256$-$1257~b nor G\,196-3~B are detected at the NOEMA 1.3-mm band. In addition, VHS\,J1256$-$1257~b remains undetected at the shorter ALMA and longer VLA wavelengths. There is no thermal emission detectable at the depth of our sub-mm, mm, and cm observations that may robustly confirm the presence of surrounding disks. We built the spectral energy distributions (SEDs) of both targets by adopting the flux upper limits with a 3-$\sigma$ confidence for all NOEMA, ALMA and VLA data. To complete the optical through mid-infrared wavelengths, we compiled all available photometry and spectroscopy from the literature (see references in \citealt{ZapateroOsorio2010,Gauza2015,Guirado2018,Climent2022}). Figure~\ref{fig:SEDs} shows the final SEDs. 

\par To compute the mass upper limits on any hypothetical dusty disk surrounding VHS\,J1256$-$1257~b and G\,196-3~B, we used the following equation \citep{Beckwith1990,Ricci2017}, which is valid for optically-thin, gas-depleted disks (i.e., we assumed that very little gas is left by the ages of our targets):
\begin{equation}
\label{Md}
  M_{\rm dust} = \frac{d^2F_{\nu}}{B_{\nu}(T_{\rm d})\kappa_{\nu}},
    \end{equation}
where $d$ is the distance to the source, $B_{\nu}(T_{\rm d})$ is the Planck function evaluated at the characteristic temperature of the emitting dust ($T_{\rm d}$), and $F_{\nu}$ is the continuum flux density at the given frequency $\nu$. The wavelength-dependent opacity $\kappa_{\nu}$ was calculated for the NOEMA and ALMA data using the code {\texttt OpacityTool} available through the Project Diana\footnote{\url{https://diana.iwf.oeaw.ac.at/data-results-downloads/fortran-package/}} introduced by \citet{2016A&A...586A.103W} and for the default dust parameters therein{\footnote{The default dust parameters in the {\texttt OpacityTool} code are the following: dust particles size distribution from 0.005~$\mu m$ to 3000~$\mu m$ with the powerlaw of -3.5, the porosity (so the volume fraction of vacuum) of 0.25, mass and volume fraction of carbonaceous material of 0.13 and 0.15, respectively.}}, except for the dust particles size distribution that we used for the range from 0.1 to 100 $\mu$m. For the VLA wavelengths, we used the extrapolation of the \cite{Beckwith1990} opacities to longer wavelengths{\footnote{$\kappa_{\nu} = 4.9$~cm$^2$/g at $\nu$ = 340~GHz ($\lambda$ = 0.88~mm),\\ $\kappa_{\nu} = 2.3$~cm$^2$/g at $\nu$ = 230~GHz ($\lambda$ = 1.3~mm),\\ $\kappa_{\nu} = 0.33$~cm$^2$/g at $\nu$ = 33~GHz ($\lambda$ = 0.9~cm),\\ $\kappa_{\nu} = 0.10$~cm$^2$/g at $\nu$ = 10~GHz ($\lambda$ = 3~cm),\\ $\kappa_{\nu} = 0.06$~cm$^2$/g at $\nu$ = 6~GHz ($\lambda$ = 5~cm),\\ $\kappa_{\nu} = 0.014$~cm$^2$/g at $\nu$ = 1.4~GHz ($\lambda$ = 21.4~cm)}}. $\kappa_{\nu}$ suffers from a large uncertainty since it depends not only on the dust physical properties (e.g. \citealt{Semenov2003, 2016A&A...586A.103W}) but on the physical processes happening inside the disk as well \citep[e.g.][]{Isella2014, Wu2022}. 

The temperature of the dust was evaluated using the relation of \citet{vanderPlas2016}: $T_{\rm d} = a\: (L_{\ast}/L_{\odot})^{b}$, where $L_{\ast}/L_{\odot}$ is the bolometric luminosity of the disk-host source in solar units, and $a$ and $b$ are coefficients dependent on the disk size{\footnote{Depending on the disk possible $R_{\rm d}$ we used the following coefficients:\\
for $R_{\rm d}<10$~AU we used [a,b] = [57,0.23],\\
for $R_{\rm d} = 10-35$~AU:  [a,b] = [41,0.22],\\
for $R_{\rm d} = 40-70$~AU:  [a,b] = [25,0.17],\\
for $R_{\rm d} > 90$~AU: [a,b] = [22,0.16].}}.
As in \cite{Shabram_2013} and \cite{Rab2019}, for our targets, we assumed the disk outer radius ($R_{\rm d}$) as that given by 1/3 of the Hill sphere radius{\footnote{The sum in denominator accounts for the fact that in some of the systems the mass of companion is non-negligible compared to the mass of the primary.}}:
\begin{equation}\label{Rd}
  R_{\rm {Hill}} = r \left(\frac{1}{3}\frac{m}{M + m}\right)^{1/3}.
\end{equation}
where $m$ and $M$ correspond to the masses of the substellar companion and the primary, and $r$ stands for their orbital separation. 

We made the general assumption that disk-host companions at separations larger than 220~au from their primaries are far enough and the disk heating from the companions is dominating over the hitting induced by the brighter primaries (i.e companion can be considered as isolated object). G\,196-3~B falls in this category, but VHS\,J1256$-$1257~b lies closer to its primary. For this "close" companion, we computed the characteristic dust temperature assuming $T^{*4}_{d} = T_d^4 + T_{ir}^4$, where $T_{ir}$ is a temperature evaluated from the heat balance equation between the host source and the dust, roughly assuming that the distance to the dust equals to the separation of the companion and that there is no disk material in between them.

\par For VHS\,J1256$-$1257~b and G\,196-3~B, we inferred sizes of the disks of $R_{d\:max}  \sim$18 and $\sim$30 au, respectively. These values are provided in Tables~\ref{tab:properties_literature} and \ref{tab:properties_computed} together with the characteristic cold temperatures of the dusty disks. To determine the mass upper limits on the putative disks around our targets, we used the observations providing the most constraining results: the NOEMA datum for G\,196-3~B and the ALMA datum for VHS\,J1256$-$1257~b. The spectral energy distributions of the cold disks containing the computed maximum dust masses are illustrated in Fig.~\ref{fig:SEDs}; the maximum mass values are listed in Table~\ref{tab:properties_computed} in units of the Earth mass.

\section{Discussion}
\label{sec:discussion}

\subsection{Comparison to other substellar companions}

\par We put our observations and results in the context of other substellar companions of similar mass with available long-wavelength observations. Over the past few years, a number of millimeter and sub-millimeter observations targeting the circumplanetary or debris disks ended-up with mostly non detections \citep{Booth2016,Wu2017,Ricci2017,Su2017,Perez2019,Matra2019,Pineda2019,Wu2020}. Only two circumplanetary disks have been confirmed around the substellar companions PDS~70~c and SR~12~c \citep{Benisty2021,Wu2022}. Table~\ref{tab:properties_literature} lists all companions with masses smaller than $\sim$20 M$_{\rm Jup}$ that have been scrutinized for their cold disks using NOEMA, VLA and/or ALMA observations and for which only flux density upper limits are available. Their ages oscillate between 1 and a few hundred Myr. We compiled the mass determinations of each source from the literature and show the the most likely values in Table~\ref{tab:properties_literature}. $\beta$\,Pic~b has a dynamical mass derived by \citet{Lacour2021}. The companion PZ Tel~B was first published as a 3.2--24.4 M$_{\rm Jup}$ companion by \citet{schmidt2014}. Its mass was later revised to 38--72 M$_{\rm Jup}$ by \citet{maire2016}, which is the value we employed in our work.

Then, we computed dust mass upper limits for all objects in Table~\ref{tab:properties_computed} following the procedure described in Section~\ref{sec:upperLimits}. 
Excluded from Tables~\ref{tab:properties_literature} and \ref{tab:properties_computed} are LkCa15~b and HD\,100546~b presented in \cite{Ricci2017} and \cite{Pineda2019}, because their substellar mass and luminosity parameters are very uncertain. Moreover, the existence of the companions is questioned in the literature. In the case of LkCa15~b, \citet{Currie2019} claimed that all previously suspected planetary companions of the LkCa15 system (LkCa15~b,c,d) are most likely signals from the disk inner structure of the primary based on new near-infrared direct imaging/spectroscopy collected with Subaru Coronagraphic Extreme Adaptive Optics  system coupled with Coronagraphic High Angular Resolution Imaging Spectrograph (CHARIS) integral field spectrograph and multi-epoch thermal infrared imaging from Keck/NIRC2. Also excluded is the V1400 Cen system \citep{Mamajek2012,Kenworthy2015,Kenworthy2020} because of the unconstrained mass of the companion (see also the Introduction section). Some of our mass upper limit estimates deviate from the literature values. GQ\,Lup~B and GSC\,6214-210~B's maximum dust masses differ from the determinations presented in \cite{Ricci2017} because we are using new ALMA data published in \cite{Wu2017}. The limits derived for PZ\,Tel~B, 51\,Eri~b, AB\,Pic~b, $\kappa$\,And~b, HR8799~b,c,d,e, $\beta$\,Pic~b, and HD\,95086~b are also different by one order of magnitude when compared to those derived by \cite{Perez2019}. The main reason is the radiative transfer model (RADMC-3D) used by \cite{Perez2019} that gives  mostly lower dust temperatures by $\sim$30~K for disk sizes in the interval 1--10~au (for more details see Section 3 of \citealt{Perez2019}).

\begin{figure*}[!ht]
\begin{center}
{\label{fig:UpperLimits_all_Mc}\includegraphics[width=0.8\textwidth]{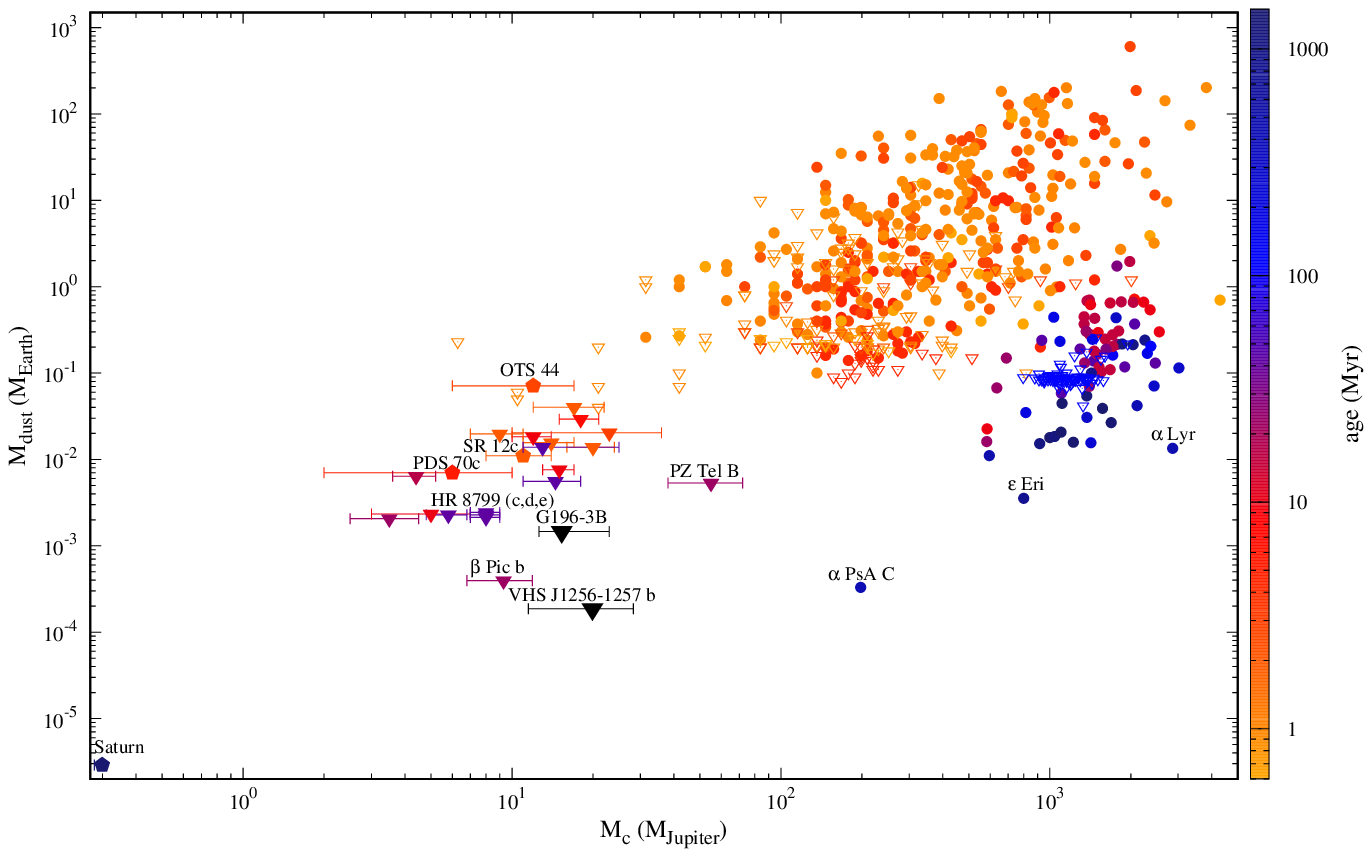}}
{\label{fig:UpperLimits_all_Mc}\includegraphics[width=0.8\textwidth]{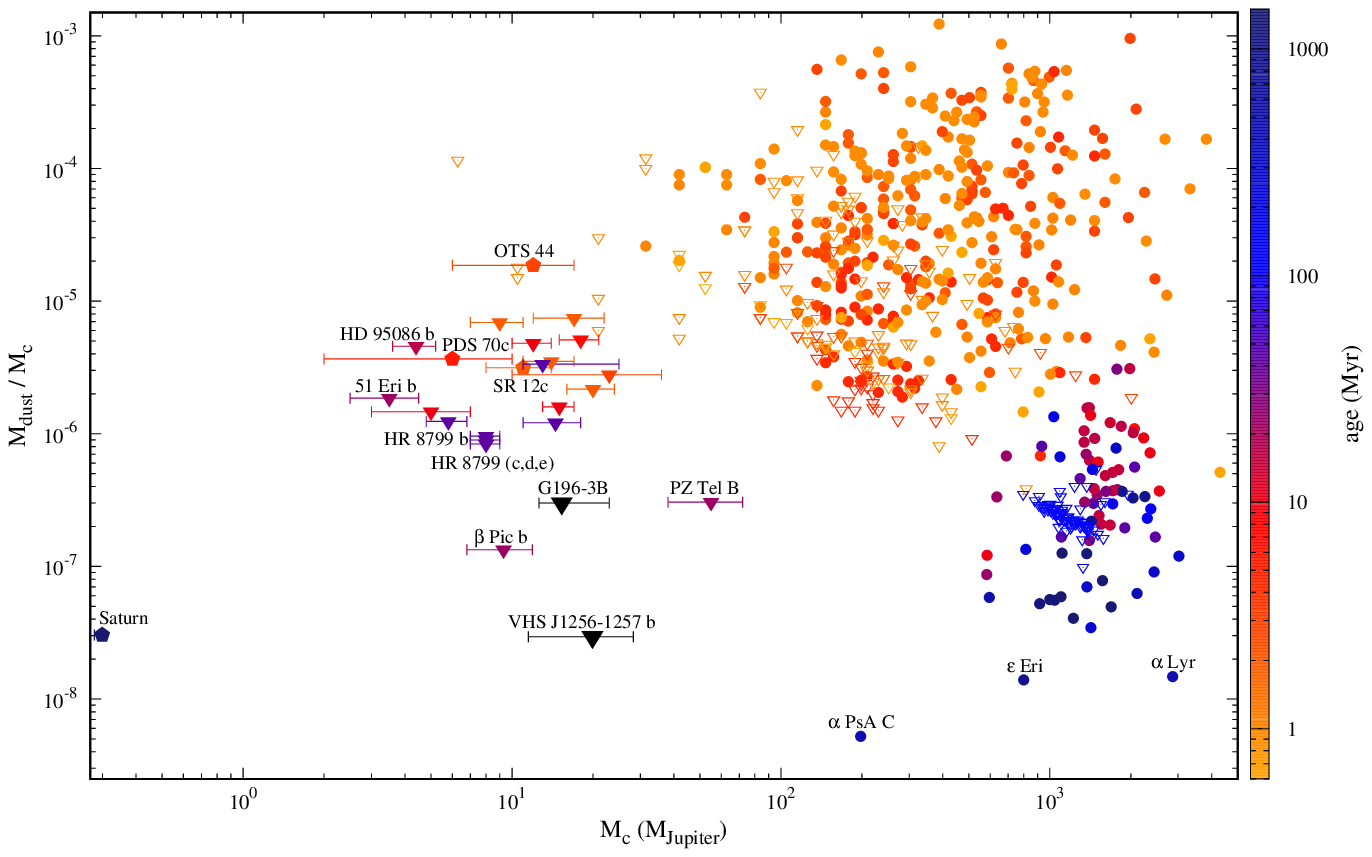}}
\caption{Measured (top panel) and relative (bottom panel) disk dust masses (pentagons and circles, ALMA detections) and dust mass upper limits (triangles, mostly ALMA non-detections) as a function of the mass of the disk-host object. Data taken from Table~2 and \citet{Testi2022}, \citet{Sepulveda2019}, \citet{Marino2020}, \citet{Sullivan2022}, and Matr\`a (2022, in preparation). Upper limits of substellar companions and free-floating objects are shown with filled and open triangles, respectively. Computations of the dust masses and upper limits are based on the assumption of a constant temperature of the disk dust (see text). In the bottom panel, the three companions of HD~8799~(c,d,e) have the same estimated dust mass upper limits; we artificially shifted them vertically so it is visually noticeable that there are 3 data points. The ages of the disk-host sources are color-coded (see the bar to the right of the panels). Protoplanetary disks (i.e., very young ages) have yellowish to orange-ish colors, while debris disks (i.e., typically older than 20 Myr) have dark red, blue, and dark blue colors, with the exception of our two targets (G\,196-3\,B and VHS\,J1256--1257\,b) plotted as black triangles.
}
\label{fig:upperLimits_all}
\end{center}
\end{figure*}

In support of the hypothesis that many of these objects can host a surrounding disk, \citet{vanHolstein2021} recently obtained linear polarimetric images in the near-infrared for 20 known directly imaged exoplanets and brown dwarf companions, 11 of which are included in Table~\ref{tab:properties_literature}: PZ~Tel~B, AB~Pic~b, GSC~6214-2010~B, GQ~Lup~B, HR~8799~(b,c,d,e), 1~RXS1609~b, DH~Tau~B, and $\beta$~Pic~b. They reported the first detection of linear polarization from substellar companions, with a polarization degree of several tenths of a percent for DH~Tau~B and GSC~6214-210~B. Given the very young ages of the two objects, the authors assumed that the polarization most likely originates from circumsubstellar protoplanetary disks. This interpretation also agrees with that of \citet{milespaez2017}, where two free-floating Taurus objects of late-M spectral types and brown dwarf masses were reported to have significant near-infrared linear polarization, in one case clearly correlated with mid-infrared flux excesses, thus consistent with the presence of a surrounding disk. For the rest of the \citet{vanHolstein2021} sample, the authors reported a null polarimetric detection but it does not necessarily mean an absence of disk. Actually, \citet{Christiaens2019} announced evidence for a disk around PDS~70~b (included in the target list of \citealt{vanHolstein2021}) based on near-infrared spectroscopic observations. 

Figure~\ref{fig:upperLimits_comanions} shows our computed upper limits on the disk dust mass as a function of the mass of the companion, which is the disk-host source. The left panel depicts the mass upper limits as measured while the right panel illustrates the mass upper limits normalized to the mass of the central companion to break any dependence of the disk dust mass on the mass or luminosity of the disk-host object. Both upper limits and the two solid detections are depicted together. As a reference, the masses of the rings of Saturn \citep{Zakhozhay2017}, the Moon and total mass of the Jovian Moons \citep{Ricci2017} are also included in the left panel. Our targets are shown with dust mass upper limits of 0.014 M$_{\rm Earth}$ (G\,196-3~B) and 0.005 M$_{\rm Earth}$ (VHS\,J1256$-$1257~b). These measurements are of the order of magnitude of the mass of the Moon.

There is a clear observational bias in both panels of Figure~\ref{fig:upperLimits_comanions}: the "older" objects ($>$10 Myr) have lower disk dust mass upper limits than the very young sources ($\le$10 Myr) because the former are typically located closer and, therefore, observations yield deeper absolute fluxes for them. This is better illustrated in Figure~\ref{fig:Md_d_age}, where the objects of Table~\ref{tab:properties_literature} are depicted as a function of their distance. G\,196-3~B, VHS\,J1256$-$1257~b (our two targets) and $\beta$~Pic~b are the closest substellar companions for which an attempt to detect surrounding disks has been made to date. The $\beta$~Pic~b observations \citep{Matra2019} are actually quite deep and our dust mass upper limit estimate is about two orders of magnitude more restrictive for $\beta$~Pic~b than for our two targets (left panel of Figure~\ref{fig:Md_d_age}). Regarding the group of younger objects (typical distance $\ge$ 100 pc), where there are two positive disk detections, both Figures~\ref{fig:upperLimits_comanions} and~\ref{fig:Md_d_age} indicate that SR~12~c and PDS~70~c may host relatively "bright" disks and that observations have to be at least one order of magnitude deeper to be able to detect similar disks around the other very young brown dwarf companions included in the Figures. 

\subsection{Comparison to free-floating sources}

For a wider view, we also compared what is known for substellar companions to the situation of free-floating brown dwarfs and stars in Figure~\ref{fig:upperLimits_all}, which shows the protoplanetary disk dust masses and dust mass upper limits using ALMA data provided by \citet{Testi2022} for the following star-forming regions and young star clusters: L1688 (in $\rho$-Ophiuchi), Taurus, Lupus, Corona Australis, Upper Scorpius, and Chameleon I. All of these regions are very young with typical ages below 10--20 Myr. For the computation of the disk dust masses around central objects with masses from substellar to several times the solar mass, \cite{Testi2022} adopted $\kappa_{\nu}$ = 2.3 cm$^{2}$\,g$^{-1}$ and a constant $T_{dust}$ = 20~K. 

For a proper comparison with the data of \cite{Testi2022}, we redid our computations of the disk dust mass upper limits for the substellar companions following the same adoption\footnote{For PDS~70~c, SR~12~c and OTS44 we also computed their disk masses adopting  $T_{dust}$~=~20~K and the equations given in this work, resulting in disk dust masses of 0.007, 0.011 and 0.071~$M_{\rm Earth}$, respectively.}  (see also \citealt{Pascucci2016}). Our new computations (provided in Table~\ref{tab:properties_computed}), \citet{Testi2022} determinations, and Saturn's rings \citep{Zakhozhay2017} are displayed in the upper panel of Figure~\ref{fig:upperLimits_all}, while the bottom panel shows the relative disk dust masses (disk mass over the mass of the central source) as a function of the disk-host object. The ages of the various objects are color coded in Figure~\ref{fig:upperLimits_all}. There are significant differences between our computations of disk dust mass upper limits and those obtained by adopting a constant dust temperature. As an example, our VHS\,J1256$-$1257~b's estimate is 15 times larger than the one from the disk constant temperature. However, this difference is still within the large dispersion for a given mass of the central object as seen in Figure~\ref{fig:upperLimits_all}. 

Similarly to stars, the detection of disks surrounding substellar objects appears to drastically depend on the systems age. The only sub-mm and mm detections in the substellar regime (OTS44, SR~12~c and PDS~70~c) in Figure~\ref{fig:upperLimits_all} correspond to sources with an age of less than 10 Myr, while there is no single disk detection around brown dwarfs with older ages despite the observations are more sensitive (due to the closer distances). This could be explained by the natural evolution of the disks, which transists from protoplanetary to the debris phase. All three OTS44, SR~12~c and PDS~70~c likely host protoplanetary disks in which gas accretion still takes place on the central objects. 

As already pointed out by many groups, the mass of the protoplanetary disks linearly depends on the mass of the central object including the substellar regime. This implies that the disks around brown dwarfs are scaled-down versions of the stellar disks. As illustrated in the bottom panel of Figure~\ref{fig:upperLimits_all}, for very young objects the relative disk dust masses appear to be constant for all stellar and substellar mass regimes, although with a relative large dispersion of about one order of magnitude. This implies that if planets form around substellar objects, the amount of dust mass available for planetary formation is small. Only rocky planets like Earth or smaller could form unless more massive planets are raised in earlier stages ($<$1 Myr) when the substellar disks might have been more massive.

Because of the age of G\,196-3~B and VHS\,J1256$-$1257~b, we would expect them to host debris disks rather than protoplanetary disks. For completeness, we also included in both panels of Figure~\ref{fig:upperLimits_all} ALMA detections of circumstellar debris disks from \citet{Sepulveda2019}, \citet{Marino2020}, and Matr\`a (2022, in preparation), and dust mass upper limits determined from the \cite{Sullivan2022} observations of Pleiades stellar members. All of these objects have typical ages greater than 20 Myr (the Pleiades age is 115 Myr, \citealt{Stauffer1998,Sullivan2022}). To determine the masses of all the central objects with detected debris disks (presented in \citealt{Sepulveda2019} and Matr\`a 2022, in preparation), we used a simple mass---luminosity relation $M_{\ast} = L_{\ast}^{0.25}$ (where both parameters are in solar units) applicable to the masses > 0.43\,M$_{\odot}$ \citep{Duric2003}\footnote{ $\alpha$\,PsA\,C is the only one object resulted to have the mass smaller than 0.43\,M$_{\odot}$, for it we recomputed the mass, using the following relation: $M_{\ast} = (L_{\ast}/0.23)^{1/2.3}$, following the prescription by \cite{Duric2003}.}. For consistency, we computed the dust masses and dust mass upper limits for all debris disks (\citealt{Sepulveda2019,Marino2020,Sullivan2022}, Matr\`a 2022, in preparation) following the approach of \citet{Testi2022}. 

In both panels of Figure~\ref{fig:upperLimits_all}, stars with debris disks are located below the stars with protoplanetary disks. As expected, the dust mass of debris disks is significantly smaller than that of protoplanetary disks for a given central source. Interestingly, the sequence described by the debris disks appears to be parallel to that of the protoplanetary disks for the central object mass interval between $\sim$0.5 and $\sim$2 M$_\odot$. Under the assumption that disks around brown dwarfs and planets of different ages behave similarly, that is, they scale down with the mass of the central body, the extrapolation of the debris disk stellar sequence into the substellar domain using the same decreasing slope observed for protoplanetary disks indicates that sub-mm and mm observations must be at least 100 times deeper than current observations in order to detect the presence of substellar debris disks surrounding moderately young brown dwarfs and planets. According to the bottom panel of Figure~\ref{fig:upperLimits_all}, only the ALMA observations of VHS J1256--1257\,b marginally comply with this requirement, yet at the upper part of the large dispersion of the predicted debris disks dust masses. That is, current observations have largely missed substellar debris disks other than ruling out their existence. These observations are indeed challenging requiring tens to hundreds of hours of observations with ALMA. Future interferometers, as SKA1-MID at high frequencies (Band~6, 15–50~GHz), with the added benefit of improved sensitivity and spatial resolution, could provide signicant contributions to the debris disks of these objects \citep{Conway2020}.

\section{Conclusions}
\label{sec:conclusions}
In this paper we presented the results of NOEMA (1.3~mm) observations of  G\,96-3~B (15.3$^{+7.7}_{-2.7}$ M$_{\rm Jup}$), and ALMA (0.87~mm), NOEMA (1.3~mm) and VLA (0.9~cm and 5~cm) observations of VHS\,J1256$-$1257~b (20.5$^{+6.0}_{-4.5}$~$M_{\rm Jup}$), both of which are low-mass companions moving around more massive objects at relatively wide orbits.  We reported non-detections for all of our observations. Based on the $rms$ noise of the millimeter observations, we determined the 3-$\sigma$ upper limits on the dust masses of any potential debris disks, and we compared them to the values (mostly obtained using ALMA observations) of other substellar companions (including brown dwarfs and planets) of different ages. Only three protoplanetary disks surrounding brown dwarfs have been detected at mm wavelengths in the literature: two around the companions PDS~70~c and SR~12~c \citep{Isella2019, Benisty2021, Wu2022}, and one around the free-floating brown dwarf OTS44 \citep{Bayo2017} despite the fact that there are at least 30 brown dwarfs and planetary mass objects with strong evidence of hosting protoplanetary disks. This demonstrates the remarkable low luminosity of these circumbsubstellar disks, which likely become much fainter when they evolve into debris disks. Our observations of G\,96-3~B and VHS\,J1256$-$1257~b  add up to the list of brown dwarf and planetary companions with non-detected debris disks. Assuming that debris disks scale down with the mass of the central bodies in a similar way as their protoplanetary predecessors do, we estimate that mm observations 100 times deeper are required for positive detections, which would open a new door for the study of the potential of these low-mass objects for generating their own moons.

\section*{Acknowledgements}
The authors are indebted to Luca Ricci, Sebastian Marino and Paola Pinilla for helpful discussions and feedback. The authors kindly thank Leonardo Tesi and Luca Matr{\`a} for providing the machine-ready tables of the protoplanetary and debris disks parameters, respectively. This work is based on observations carried out under project number W18BP [001-002] with the IRAM NOEMA Interferometer [30m telescope]. IRAM is supported by INSU/CNRS (France), MPG (Germany) and IGN (Spain). OZ acknowledges support  within the framework of the Ukraine aid package for individual grants of the Max-Planck Society 2022. MRZO acknowledges financial support from the Spanish Ministery for Science and Innovation through project PID2019-109522GB-C51. VB and NL acknowledge financial support from the Agencia Estatal de Investigaci\'on del Ministerio de Ciencia e Innovaci\'on (AEI-MCINN) under grant PID2019-109522GB-C53\@. BG is supported by the Polish National Science Center (NCN) under SONATA grant No. 2021/43/D/ST9/01940. JBC and JCG were supported by projects PID2020-117404GB-C22, funded by MCIN/AEI, PROMETEO/2020-080, funded by the Generalitat Valenciana, and by the Astrophysics and High Energy Physics programme by MCIN, with funding from European Union NextGenerationEU (PRTR-C17.I1) and the Generalitat Valenciana through grant ASFAE/2022/018. MPT acknowledges financial support through grants CEX2021-001131-S, PID2020-117404GB-C21 and PID2020-114461GB-I00, funded by the Spanish CIN/AEI/10.13039/501100011033. We also wish to thank the anonymous referee for constructive criticism that helped to improve the clarity of the paper.

\bibliographystyle{aa}
\bibliography{circumplanetaryDisks}

\clearpage
\onecolumn

\begin{landscape}
\begin{table}[htb]
\caption{Physical properties and information of mm and sub-mm observations of brown dwarfs/planetary-mass objects with non detected disks.}             
\label{tab:properties_literature}
\centering          
\begin{tabular}{lcccccccl}     
\hline\hline       
Name  & $d$\tablefootmark{a}, pc &	age, Myr &	$L_{\rm c}$\tablefootmark{a}, L$_{\odot}\times10^{-3}$ &	$M_{\rm c}$, M$_{\rm Jup}$ &	$a$, AU  & $F_{\nu}$ (3$\sigma$), mJy & $F_{\nu}$ observed with &  Refs.\tablefootmark{b}\\
\hline
G\,196-3~B	& 21.81$\pm0.01$ & 50--100 & 0.0927$\pm0.0001$ & 15.3$^{+7.7}_{-2.7}$ & $\sim$350\tablefootmark{c} & 0.108 & NOEMA (1.3 mm)  & this work \\
VHS\,J1256$-$1257~b & 21.15$\pm0.21$ & 150--300 & 0.027$\pm0.0001$\tablefootmark{d} & 20.5$^{+6.0}_{-4.5}$ & $\sim$170\tablefootmark{c} & 0.153 & NOEMA (1.3 mm)  & this work \\
&&&&&& 0.060 & ALMA 0.87 mm)  & this work \\
&&&&&& 0.030 & VLA (0.9 cm)  & this work\tablefootmark{e},  1 \\
&&&&&& 0.009 & VLA (3 cm)  & this work\tablefootmark{e}, 1, 2 \\
&&&&&& 0.012 & VLA (5 cm)    & this work\tablefootmark{e}, 1 \\
&&&&&& 0.024 & VLA (21.4 cm) & this work\tablefootmark{e}, 1, 2 \\
CT~Cha~B	& 190.0$^{+0.4}_{-0.5}$    & 1--3   & 2.38                   & 17$\pm$5            & $\sim$511            & 0.156 & ALMA(0.88 mm) & 3 \\
1~RXS1609~b	& 138.0$^{+0.4}_{-0.3}$    & 5--14  & 0.41                   & 12$\pm$2            & $\sim$309            & 0.135 & ALMA(0.88 mm) & 3 \\
ROXs~12~B 	& 138.6$^{+0.3}_{-0.4}$    & 4--10  & 1.75                   & 18$\pm$3            & $\sim$243            & 0.213 & ALMA(0.88 mm) & 3 \\
ROXs~42B~b	& 146.4$^{+0.7}_{-0.6}$    & 1--3   & 0.83                   & 9$\pm$2             & $\sim$168            & 0.129 & ALMA(0.88 mm) & 3 \\
DH~Tau~B	& 133.4$^{+0.6}_{-0.4}$    & 1--3   & 1.82                   & 14$\pm$3            & $\sim$317            & 0.123 & ALMA(0.88 mm) & 3 \\
FU~Tau~B	& 128.6$^{+1.4}_{-1.2}$    & 1--3   & 3.13                   & 20$\pm$4            & $\sim$747            & 0.117 & ALMA(0.88 mm) & 3 \\
PZ~Tel~B	& 47.25$\pm$0.05           & 20--25 & 2.6$\pm$0.6            & 38--72              & $\sim$23             & 0.084 & ALMA(1.3 mm)  & 4, 5, 6 \\
51~Eri~b	& 29.91$\pm$0.07           & 20--25 & 3$^{+2}_{-1}\times$10$^{-3}$ & 2.5--4.5      & 11.1$^{+4.2}_{-1.3}$  & 0.081 & ALMA(1.3 mm)   & 4, 7, 8 \\
AB~Pic~b	& 50.14$\pm$0.03           & 30--55 & 0.22$^{+0.05}_{-0.04}$ & 11--18              & $\sim$252            & 0.078 & ALMA(1.3 mm)  & 4, 9 \\
$\kappa$~And~b	& 52.0$\pm$0.5             & 40--50 & 0.17$^{+0.02}_{-0.02}$      & 13$^{+12}_{-2}$     & $\sim$55              & 0.18  & ALMA(1.3 mm)  & 4, 10  \\
$\beta$~Pic~b	& 19.63$^{+0.06}_{- 0.05}$ & 20--25 & 0.18$^{+0.01}_{-0.01}$      & 9.3$^{+2.6}_{-2.5}$ & 10.26$^{+0.1}_{-0.1}$       & 0.036 & ALMA(1.3 mm)   & 4, 11-13 \\
GSC~6214-2010~B	& 108.78$\pm$0.25          & 5--14      & 0.447          & 15$\pm$2            & 320                  & 0.09  & ALMA(0.88 mm)  & 14, 15 \\
GQ~Lup~B	& 154.1$\pm$0.7            & 3$\pm$2    & 3.306              & 10--36              & $\sim$220            & 0.12  & ALMA(0.88 mm)  & 14, 16 \\
HD~95086~b	& 86.46${\pm}$0.14         & 17$\pm$4  & 0.017              & 4.4${\pm}$0.8	   & 57.2$^{+19.2}_{-7.8}$ & 0.03  & ALMA(1.3 mm)   & 4, 17 \\
2M~1207~b	& 64.7$\pm$0.5           & 10$\pm$3     & 0.225              & 5$\pm$2             & $\sim$40              & 0.078 & ALMA(0.89 mm)  & 18 \\
HR~8799~b	    & 40.88$\pm$0.08  & 42$^{+6}_{-4}$     & 0.007           & 5.8                 & 70                    & 0.048 & ALMA(1.3 mm)  & 4, 19 \\
HR~8799~c	& 40.88$\pm$0.08  & 42$^{+6}_{-4}$     & 0.013           & 7--9                & 43            & 0.048 & ALMA(1.3 mm)   & 4, 19 \\
HR~8799~d	& 40.88$\pm$0.08  & 42$^{+6}_{-4}$     & 0.013           & 7--9                & 26            & 0.048 & ALMA(1.3 mm)   & 4, 19 \\
HR~8799~e	& 40.88$\pm$0.08  & 42$^{+6}_{-4}$     & 0.013           & 7--9                & 15            & 0.048 & ALMA(1.3 mm)   & 4, 19 \\
\hline                  
\end{tabular}
\tablefoot{
\tablefoottext{a}{All the distances are computed from the GAIA DR3~\citep{GaiaCollaboration2016,Lindegren2021} and the luminocities are updated accordingly.}
\tablefoottext{b}{References for the observed fluxes and systems physical parameters.}
\tablefoottext{c}{The separation is computed using the latest GAIA DR3 distance.}
\tablefoottext{d}{The luminosity of VHS\,J1256$-$1257~b was obtained by integrating the photometric observed fluxes (blue dots in figure~\ref{fig:SEDs}) over over the extended wavelength range 0.3–100$\mu$m (Simpson’s rule). For more details see eq.1 in~\cite{Zakhozhay2017}.}
\tablefoottext{e}{For more details see \cite{Climent2022}.}
}
\end{table}

\tablebib{
(1)~\mbox{\citet{Guirado2018}}
(2)~\mbox{\citet{Climent2022}}
(3)~\mbox{\citet{Wu2020}};
(4)~\mbox{\citet{Perez2019}};
(5)~\mbox{\citet{Miret-Roig2018}};
(6)~\mbox{\citet{Maire2018}};
(7)~\mbox{\citet{Dupuy2022}};
(8)~\mbox{\citet{DaRosa2020}};
(9)~\mbox{\citet{Neuhauser2012}};
(10)~\mbox{\citet{Currie2018}};
(11)~\mbox{\citet{Brandt2021}};
(12)~\mbox{\citet{Chilcote2017}};
(13)~\mbox{\citet{Matra2019}};
(14)~\mbox{\citet{Wu2017}};
(15)~\mbox{\citet{Bowler2015}};
(16)~\mbox{\citet{MacGregor2017}};
(17)~\mbox{\citet{Su2017}};
(18)~\mbox{\citet{Ricci2017}};
(19)~\mbox{\citet{Booth2016}};
}
\end{landscape}

\begin{table}[htb]
\caption{Radius and mass upper limits and temperature of potential dusty disks around brown dwarf and planets with non detections.}            
\label{tab:properties_computed}
\centering          
\begin{tabular}{lrrrrl}     
\hline\hline       
Name  &	$R_{\rm out}$\tablefootmark{a} &	$T_{\rm d}$  &	$M_{\rm dust}$\tablefootmark{b} &	$M_{\rm dust,20K}$\tablefootmark{c} &  $F_{\nu}$\tablefootmark{d} observed with\\
 & (au) & (K) & (M$_{\oplus}$) & (M$_{\oplus}$) &  \\
\hline
G\,196-3~B	& 29.5 & 5.3  & 0.014 & 0.0015 & NOEMA (1.3 mm) \\
VHS\,J1256$-$1257~b & 18.1 & 4.5 &  0.028 & 0.0019 & NOEMA (1.3 mm) \\
                                &&&  0.005 & 0.00019 & ALMA 0.87 mm) \\
                                &&&  0.500 & 0.098 & VLA (0.9 cm) \\
                                &&&  4.889 & 1.06 & VLA (3 cm) \\
                                &&&  29.535 & 6.5 & VLA (5 cm) \\
                                &&&  4519.046 & 1015.1 & VLA (21.4 cm) \\
CT~Cha~B	&  32.0      &  10.9  &  0.111 & 0.040 & ALMA(0.88 mm) \\
1~RXS1609~b	&  16.9     &  7.4   &  0.118 & 0.018 & ALMA(0.88 mm) \\
ROXs~12~B 	&  15.1     &  10.1  &  0.092 & 0.029 & ALMA(0.88 mm) \\
ROXs~42B~b	&  8.2     &  11.1  &  0.052 & 0.020 & ALMA(0.88 mm) \\
DH~Tau~B	&  21.7     &  11.7 &  0.037 & 0.016 & ALMA(0.88 mm) \\
FU~Tau~B	&  101.0     &  8.7  &  0.060 & 0.014 & ALMA(0.88 mm) \\
PZ~Tel~B	&  2.0     & 60.7   &  0.001 & 0.0053 & ALMA(1.3 mm) \\
51~Eri~b	&  0.3     & 128.5  &  3$\times10^{-4}$ & 0.0021 & ALMA(1.3 mm) \\
AB~Pic~b	&  15.0     & 6.4    &  0.035 & 0.0056 & ALMA(1.3 mm) \\
$\kappa$~And~b	& 2.2  & 113.4  &  0.002 & 0.014 & ALMA(1.3 mm)8  \\
$\beta$~Pic~b	&  0.4  & 152.1  &  4$\times10^{-5}$ & 0.00039 & ALMA(1.3 mm) \\
GSC~6214-2010~B	& 18.5  & 7.5    &  0.047 & 0.0076 & ALMA(0.88 mm)\\
GQ~Lup~B	&  13.9     & 18.9   &  0.022 & 0.020 & ALMA(0.88 mm) \\
HD~95086~b	&  1.8      & 60.9   &  0.002 & 0.0064 & ALMA(1.3 mm) \\
2M~1207~b	&  5.1     & 12.1   &  0.005 & 0.0023 & ALMA(0.89 mm) \\
HR~8799~b	&  2.5      & 51.2   &  7$\times10^{-4}$ & 0.0023 & ALMA(1.3 mm) \\
HR~8799~c	&  1.7      & 65.3   &  6$\times10^{-4}$ & 0.0023 & ALMA(1.3 mm) \\
HR~8799~d	&  1.0      & 84.0   &  4$\times10^{-4}$ & 0.0023 & ALMA(1.3 mm) \\
HR~8799~e	&  0.6      & 110.6  &  3$\times10^{-4}$ & 0.0023 & ALMA(1.3 mm) \\
\hline                  
\end{tabular}
\tablefoot{
\tablefoottext{a}{Dust disk radii derived in this work as described in Section~\ref{sec:upperLimits}}.
\tablefoottext{b}{Dust masses computed using Equation~\ref{Md} and $T_{\rm d}$, computed using the aproach described in Section~\ref{sec:upperLimits} and listed here.}
\tablefoottext{c}{Dust masses computed using Equation~\ref{Md} and a constant $T_{\rm d}$\,=\,20\,K as in \citet{Testi2022}.}
\tablefoottext{d}{The instruments employed for collecting the information on $F_\nu$ used for the calculation of $M_{\rm dust}$.}
}
\end{table}

\begin{appendix}
\section{NOEMA and ALMA images for G\,196-3~B and VHS\,J1256-1257\,b.}
\label{sec:images}
\begin{figure}[!ht]
\begin{center}
\subfloat[]{\includegraphics[width=0.315\textwidth]{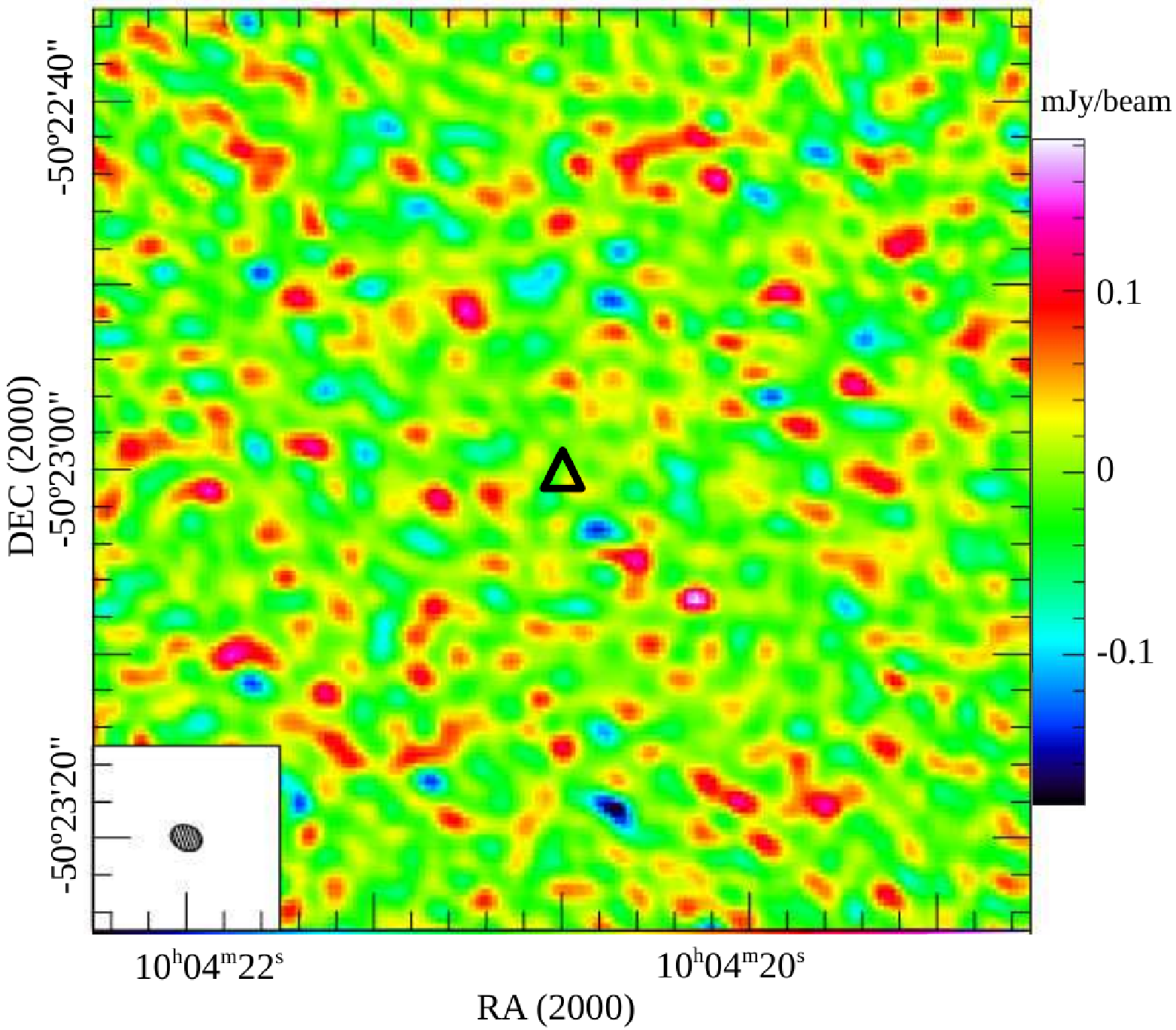}}
\hspace{0.15cm}
\subfloat[]{\includegraphics[width=0.31\textwidth]{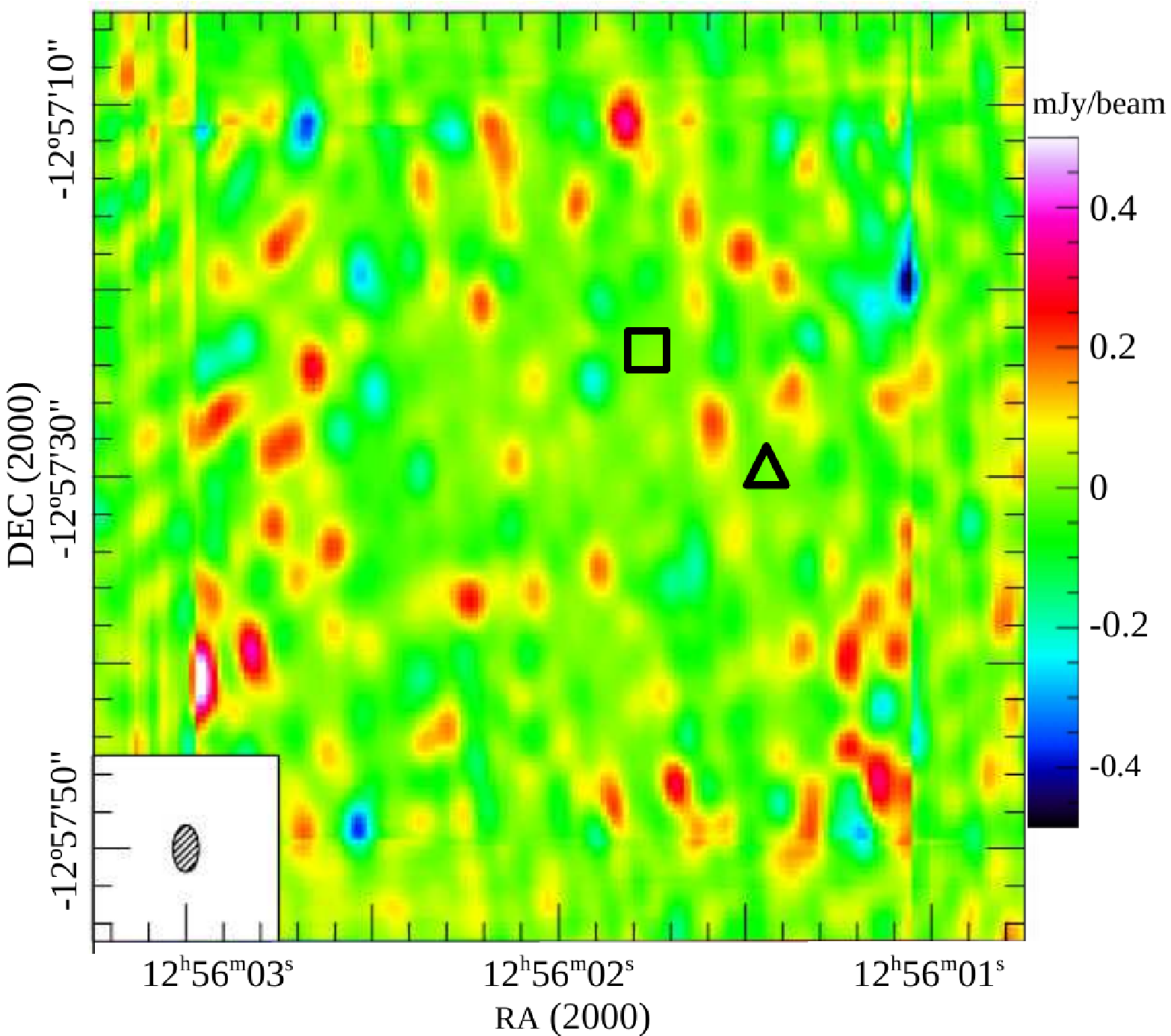}} 
\hspace{0.15cm}
\subfloat[]{\includegraphics[width=0.34\textwidth]{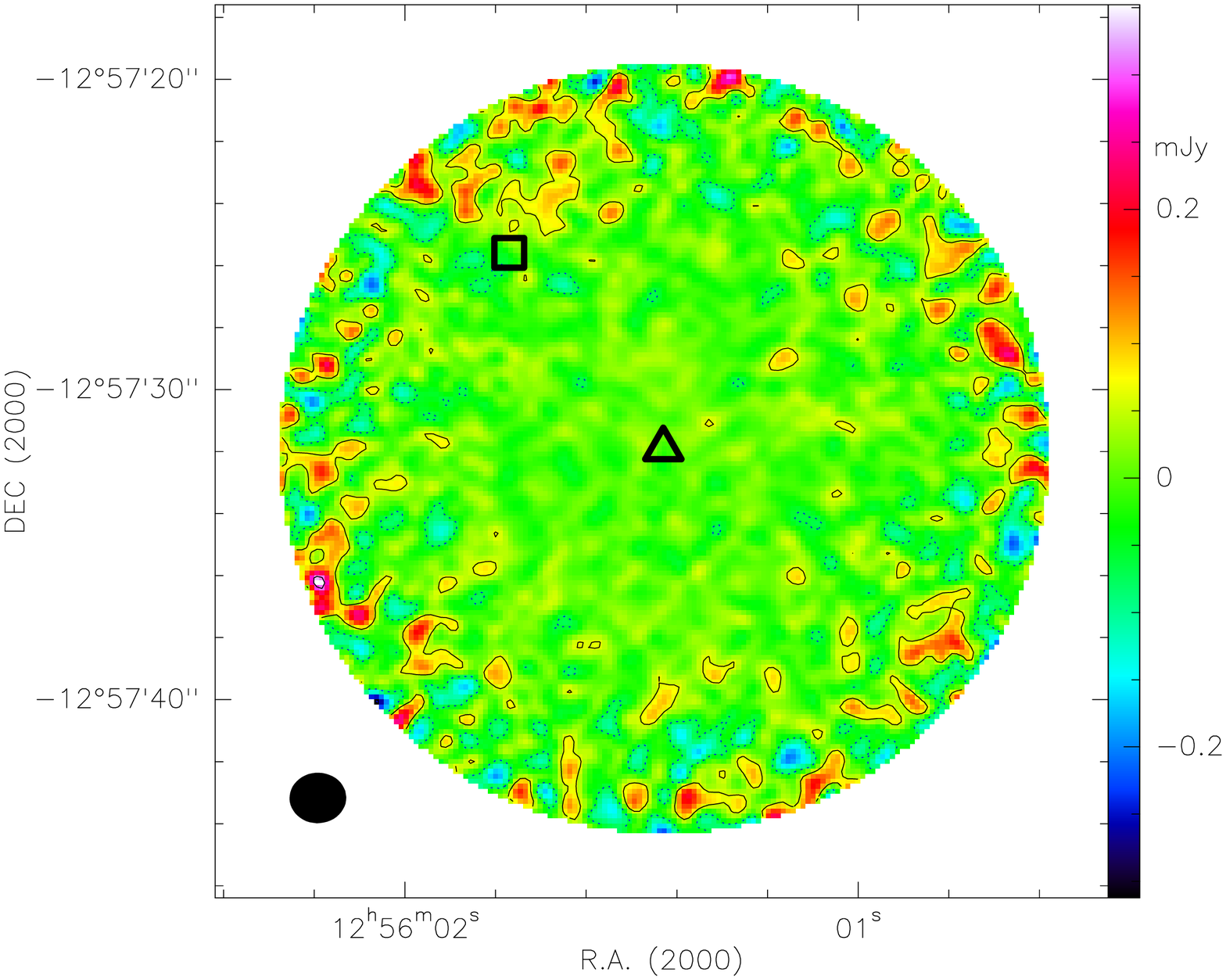}}
\caption{
NOEMA image of G\,196-3~B (a) and images VHS\,J1256-1257 with NOEMA (b) and ALMA (c). The positions of VHS\,J1256-1257\,b and G\,196-3~B are indicated with triangles, respectively. Additionally position of VHS\,J1256-1257\,AB is also indicated with square at (b) and (c) panels. The synthesized beam at (a) and (b) is indicated by the ellipse in the white square at the lower left corner.  The beam size at (c) is shown by the black quasi-circular symbol. Flux level in mJy/beam is color-coded at (a) and (b), and in mJy is color-coded at (c).
}
\label{fig:maps}
\end{center}
\end{figure}

Figure~\ref{fig:maps} shows  the NOEMA images of G\,196-3~B and VHS\,J1256-1257, and ALMA image of VHS\,J1256-1257. No circumstellar dust emission is detected from any of these images.

\end{appendix}
\end{document}